%% file: main.tex
\documentclass[10pt,conference,final]{IEEEtran}
\IEEEoverridecommandlockouts
\pdfoutput=1 

\usepackage{fancyhdr}
\usepackage{tcolorbox}
\usepackage{dirtytalk}
\usepackage{comment}
\usepackage{booktabs}
\usepackage{array}
\usepackage{cite}
\usepackage{amsmath,amssymb,amsfonts}
\usepackage{graphicx}
\usepackage{textcomp}
\usepackage{xcolor}
\usepackage{float}
\usepackage{balance}
\usepackage{hyperref}
\usepackage[normalem]{ulem}
\usepackage[linesnumbered,ruled,vlined]{algorithm2e}

\SetCommentSty{mycommfont}
\usepackage[a-1b]{pdfx}

\usepackage[utf8]{inputenc}
\usepackage{pgfplots}
\DeclareUnicodeCharacter{2212}{−}
\usepgfplotslibrary{groupplots,dateplot}
\usetikzlibrary{patterns,shapes.arrows}
\pgfplotsset{compat=newest}

\def\BibTeX{{\rm B\kern-.05em{\sc i\kern-.025em b}\kern-.08em
    T\kern-.1667em\lower.7ex\hbox{E}\kern-.125emX}}

\input{macros}

\include{data/data}
\begin{document}

\pagestyle{fancy}
\fancyhf{}
\fancyhead[C]{Published in 2023 IEEE/ACM 45th International Conference on Software Engineering (ICSE)}

\title{FedDebug: Systematic Debugging for Federated Learning Applications \thanks{Published in 2023 IEEE/ACM 45th International Conference on Software Engineering (ICSE), pages 456-789. DOI: \href{https://ieeexplore.ieee.org/document/10172839}{\color{blue} 10.1109/ICSE48619.2023.00053}
}}

\author{\IEEEauthorblockN{Waris Gill}
\IEEEauthorblockA{\textit{Computer Science Department} \\
\textit{Virginia Tech}\\
Blacksburg, USA \\
waris@vt.edu}
\and
\IEEEauthorblockN{Ali Anwar}
\IEEEauthorblockA{\textit{Computer Science and Engineering Department} \\
\textit{University of Minnesota Twin Cities}\\
Minneapolis, USA \\
aanwar@umn.edu}
\and
\IEEEauthorblockN{Muhammad Ali Gulzar}
\IEEEauthorblockA{\textit{Computer Science Department} \\
\textit{Virginia Tech}\\
Blacksburg, USA \\
gulzar@cs.vt.edu}
\and
}

\maketitle

\begin{abstract}
 \input{paper_sections/section_abstract.tex}

\end{abstract}


\input{paper_sections/section_introduction}
\input{paper_sections/section_background}


\input{paper_sections/section_debugging-primitives}

\input{paper_sections/section_fault_localization.tex}

\input{paper_sections/section_evaluation}

\input{paper_sections/section_threat_to_validity.tex}

\input{paper_sections/section_related-work}

\input{paper_sections/section_conclusion}

\bibliographystyle{plain.bst}

\bibliography{main}

\end{document}

%% file: macros.tex
\usepackage{xspace}

\usepackage{pifont}
\usepackage{pgfplots}
\usepackage{subcaption}
\usepackage{array}
\newcolumntype{L}[1]{>{\raggedright\let\newline\\\arraybackslash\hspace{0pt}}m{#1}}
\newcolumntype{C}[1]{>{\centering\let\newline\\\arraybackslash\hspace{0pt}}m{#1}}
\newcolumntype{R}[1]{>{\raggedleft\let\newline\\\arraybackslash\hspace{0pt}}m{#1}}

\usepackage[all]{nowidow}

\usetikzlibrary{patterns}
\usetikzlibrary{intersections}
\usetikzlibrary{calc}
\usetikzlibrary{arrows.meta}
\usetikzlibrary{shapes.misc}

\newcommand{\ie}{\emph{i.e.,}\xspace}
\newcommand{\eg}{\emph{e.g.,}\xspace}

\newcommand{\tool}{\textsc{FedDebug}\xspace}

%% file: paper_sections/section_abstract.tex
In Federated Learning (FL), clients independently train local models and share them with a central aggregator to build a global model. Impermissibility to access clients' data and collaborative training make FL appealing for applications with data-privacy concerns, such as medical imaging. However, these FL characteristics pose unprecedented challenges for debugging. When a global model's performance deteriorates, identifying the responsible rounds and clients is a major pain point. Developers resort to trial-and-error debugging with subsets of clients, hoping to increase the global model's accuracy or let future FL rounds retune the model, which are time-consuming and costly.

We design a systematic fault localization framework, \tool, that advances the FL debugging on two novel fronts. First, \tool enables interactive debugging of realtime collaborative training in FL by leveraging record and replay techniques to construct a simulation that mirrors live FL. \tool's {\em breakpoint} can help inspect an FL state (round, client, and global model) and move between rounds and clients' models seamlessly, enabling a fine-grained step-by-step inspection. Second, \tool automatically identifies the client(s) responsible for lowering the global model's performance without any testing data and labels---both are essential for existing debugging techniques. \tool's strengths come from adapting differential testing in conjunction with neuron activations to determine the client(s) deviating from normal behavior. \tool achieves 100\% accuracy in finding a single faulty client and 90.3\% accuracy in finding multiple faulty clients. \tool's interactive debugging incurs 1.2\% overhead during training, while it localizes a faulty client in only 2.1\% of a round's training time.  With \tool, we bring effective debugging practices to federated learning, improving the quality and productivity of FL application developers.

%% file: paper_sections/section_introduction.tex
\section{Introduction}

Many machine learning models today require private user information for high-quality training. However, users are naturally reluctant to share such data to minimize the risk of privacy violation. 
To address the above needs, Federated Learning (FL) \cite{mcmahan2017communication} enables individual participating clients (\eg smart-home edge devices) to train a machine learning (ML) model on their local data in a privacy-preserving environment and then send the trained model (\eg~the weights of the neural network) to a central aggregator to build a global model. FL trains highly accurate models without ever accessing user data, keeping clients' data privacy intact~\cite{kairouz2021advances}. With the advent of frameworks like FedML \cite{Fedml} and IBMFL \cite{ibmfl2020ibm}, FL is actively used in solving real-world problems~\cite{jiang2020federated, rieke2020future, long2020federated, zheng2021applications}.

\noindent{\textit{\textbf{Problems.}}} The support for collaborative yet privacy-preserving training in FL comes at the cost of transparency and comprehension, making debugging prohibitively complicated. For instance, a faulty client can send an inaccurate model to the aggregator either due to noisy labels~\cite{hendrycks2018using, li2019learning, li2020dividemix} 
in the training data or malicious intent to deteriorate the global model's performance \cite{biggio2012poisoning, bhagoji2019analyzing, bagdasaryan2020backdoor, ozdayi2021defending}. Finding such a faulty client is challenging due to a large number of unpredictable clients that may not have participated in every round because of a poor network connection or low battery power ~\cite{xu2019exploring, tang2021battery}. The FL training process also spans numerous rounds, significantly increasing 
the search space for identifying the true culprit round. None of the existing FL frameworks provide debugging and testing support to developers when building FL applications~\cite{kairouz2021advances}. 
These developers rely on guesswork and expensive trial-and-error debugging to find a fault-inducing client.


\noindent{\textit{\textbf{Challenges.}}} FL poses two fundamental challenges when designing a debugging technique. First, in FL deployments, training and testing data are kept private and strictly reside with clients. Access to such data could allow developers to evaluate individual clients' models sent to the aggregator and identify the lowest-performing model as the culprit, similar to traditional ML model testing. Neither test data nor labels are available to an FL application developer and, therefore, existing ML debugging approaches~\cite{pei2017deepxplore, odena2019tensorfuzz, wardat2021deeplocalize} are inapplicable. 

Second, due to the unpredictability of clients' participation in a round and the ephemeral nature of their contributions in the global model, reproducing a fault (\ie faulty client) and then debugging it is not feasible. Traditional breakpoint debugging will pause the entire training process in FL across all clients, causing severe side effects such as data loss as clients may not have persistent storage to store data. Live postmortem or trial-error debugging  may lead to a new set of clients for each round based on client availability and quorum, thus making debugging even more ineffective. Considering the above limitations and challenges, we must design a debugging approach that does not rely on clients' data, can debug a live FL application without any interference, and can localize a faulty client precisely.

\noindent{\textit{\textbf{Contributions.}}} We take inspiration from traditional debuggers, such as {\tt gdb}, and redesign traditional debugging constructs that are tailored to the needs of an FL application developer. Our approach, \tool, selectively records an FL application's telemetry data to enable realtime interactive debugging on a simulation that mirrors a live FL application. With \tool's {\em breakpoint}, a developer can spawn a simulation of a live FL application and inspect the current state containing information such as clients' models and their reported metrics (\eg their training loss or hyperparameters). 
It also allows a seamless transition between the rounds and clients at a given breakpoint, enabling a fine-grained step-by-step inspection of the application's state. When a developer finds a suspicious state (\eg multiple clients report high training loss), \tool's automated fault localization approach  precisely identifies the faulty client(s) without any test data or labels. Once a faulty client is identified, \tool's {\em fix and replay} repairs the global training  by retroactively removing the faulty client and resuming the live FL training.



\noindent{\textit{\textbf{Key Insights.}}} \tool leverages several insights to enable systematic FL debugging while preserving clients' privacy. We observe that instead of debugging a live FL application, we can record a set of runtime metrics essential to regenerate a given state in an FL application. Thus, \tool performs debugging on a regenerated simulated state equivalent to a live state. 
To have a measurable impact on the global model, a faulty client's model must behave differently than the regular clients. Every client in an FL application has the same model architecture, so their internal behaviors are comparable. Based on this insight, \tool proposes an inference-guided test selection method to select high-quality and diverse test data from a pool of randomly generated input images using Kaiming Initialization~\cite{he2015delving}.
However, an auto-generated data does not include the class label \ie an oracle. To address the {\em oracle} problem with such data, \tool adapts differential testing to FL domain. It captures differences in the models' execution via neuron activations instead of output labels to identify \emph{diverging} behavior of faulty clients. 



\noindent{\textit{\textbf{Evaluations.}}} We perform large-scale, extensive evaluation of \tool on popular models, two large-scale datasets, two well-established FL data distributions, and a real-world fault-injection technique in a total of 68 different FL configurations. 
We measure \tool's fault localizability, debugging time, performance overhead over a vanilla FL framework (IBMFL), and scalability. \tool shows remarkable success in identifying faulty clients. It can localize the real-world faulty client with 100\% accuracy within 2.1\% of a round's training time. When faced with multiple faulty clients, \tool retains the high fault localization accuracy of 90.3\%. \tool's debugging constructs incur an overhead of 48\% of the aggregation time to record telemetry data for state regeneration. Surprisingly, this time is only 1.2\% of a single round's training time in our experiments. Through our evaluation, we demonstrate that \tool effectively conducts interactive debugging and efficiently automates fault localization without incurring high runtime costs. \tool augments the IBMFL framework, but its underlying insights can be adapted for other FL frameworks. 

We summarize \tool's contributions below:
    \begin{itemize}
        \item \textbf{Originality:} To the best of our knowledge, \tool is the first general-purpose debugging framework for federated learning applications that is not limited by access to clients' data. It addresses open debugging challenges in FL \cite{kairouz2021advances}.
        \item \textbf{Approach:} Traditional ML trains a single model, whereas FL involves distributed training across hundreds of clients over multiple rounds. Thus, existing ML debugging approaches are inapplicable on FL. \tool's novelty lies in observations about FL and the exploitation of insights on  reproducibility, inference guided test generation, and differential testing that do not impede performance or violate FL privacy principles.
        \item \textbf{Benchmark:} We evaluate \tool in 68 FL configurations derived from well-established datasets, models, varying clients, data distribution, and fault-injections. We package our experiment environment into a public benchmark for future research use. 
        \item \textbf{Usefulness:} Our extensive experiments demonstrate that \tool successfully locates faulty client(s) without impeding the FL workflow. On a wide range of experiments, \tool exhibits robust results against multiple faulty clients, challenging data distributions, and a large number of clients. \tool's artifact  and the benchmarks used in this work are  publicly available at {\color{blue} \url{ https://github.com/SEED-VT/FedDebug}}.
    \end{itemize}

%% file: paper_sections/section_background.tex
\section{Background and Motivation} 
  \begin{figure}[t]
  \centering
       \includegraphics[width=0.45\textwidth]{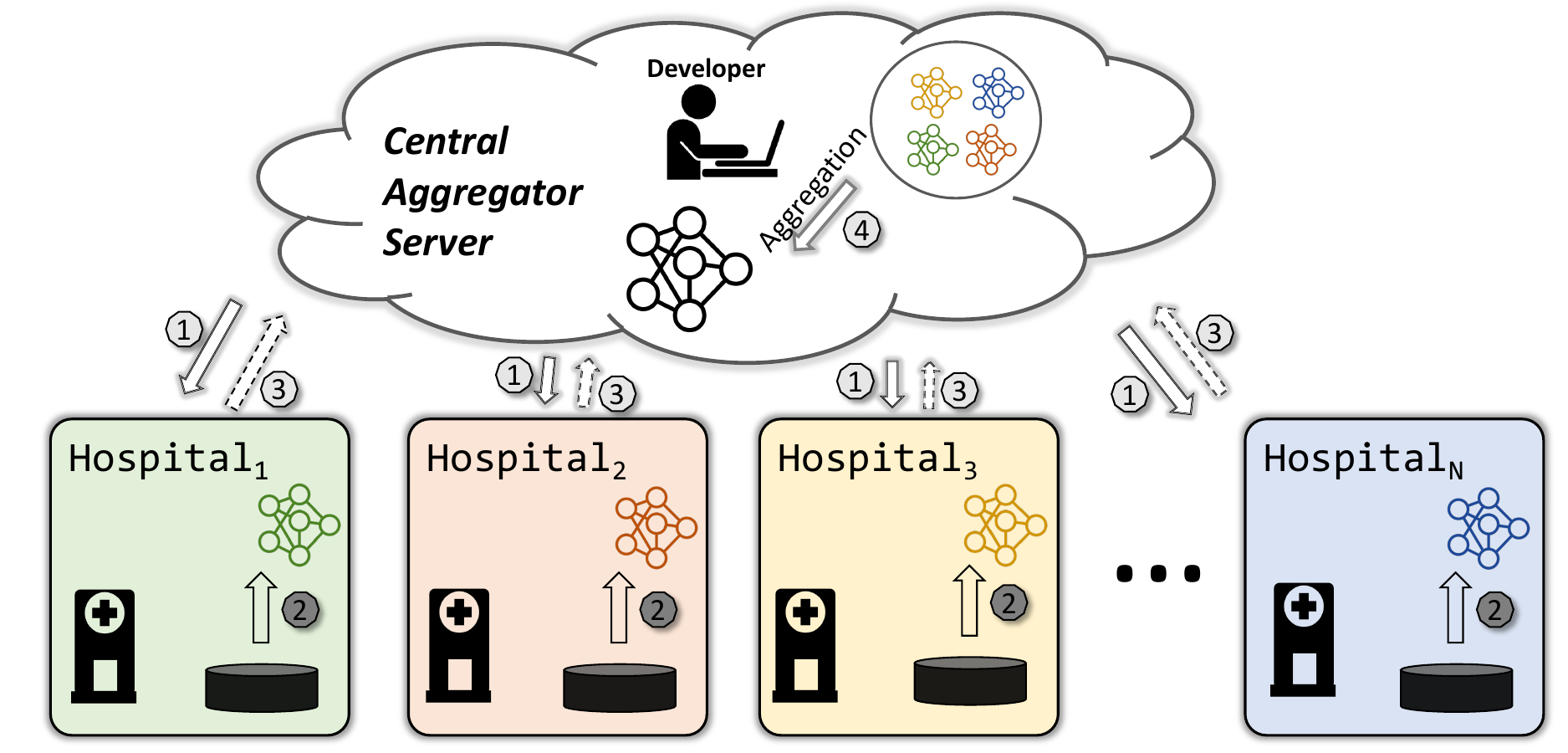}
        \caption{In a centralized FL architecture, an aggregator sends a global model to clients (step 1). Each client trains the model on local data (step 2) and sends the locally trained model back to the server (step 3). The server aggregates all models to form a new global model (step 4).}
        \label{fig:fl_base}
        \vspace{-2ex}
    \end{figure}

\subsection{Federated Learning}
\label{background}
    
    In Federated Learning, multiple clients independently train local models on their data and share it with a central server (also called an aggregator) to construct a global model. During this collaborative training, clients' training data never leaves their premises \cite{kairouz2021advances}.  
    Figure~\ref{fig:fl_base} shows an FL setting where multiple hospitals collaboratively train a global model on their local labeled medical imaging data.
    \begin{enumerate}
        \item In the first step, the aggregator sends copies of the current global model, \ie the global model weights, and hyperparameters (\eg learning rate and epochs) to participating clients (Step 1 of Figure~\ref{fig:fl_base}). 

        \item Using the global model as initial parameters, each client trains a model on its local data similar to  traditional ML training (Step 2 of Figure~\ref{fig:fl_base}). 
        
        \item Once trained, each client sends its local model, in the form of updated weights, back to the aggregator as shown in Step 3 of Figure~\ref{fig:fl_base}. Additionally, clients share  performance metrics such as training loss and quality/quantity of training data with the central aggregator. 
        \item After receiving model updates, the server aggregates the updated weights from all clients using established model aggregations (also called fusion) techniques such as FedAvg~\cite{mcmahan2017communication} to form a new global model (Step 4).
        
    \end{enumerate}

The four steps are repeated for a fixed number of {\em rounds} or until the global model meets some convergence criteria, for example,  when training loss is close to zero. Note that not every client participates in every round. There are additional variants of federated learning (FL) such as vertical FL~\cite{liu2019communication}, FL with differential privacy~\cite{wei2020federated}, and personalized FL \cite{t2020personalized}. Our work mainly focuses on the standard FL~\cite{mcmahan2017communication}.  

\subsection{Motivating Scenario}
\label{motivating}
Suppose that an FL application developer trains a global neural network model, ResNet \cite{he2015deep}, on chest X-ray images from hospitals across the country to diagnose respiratory diseases (\eg Covid-19). The term {\em developer} refers to a person who writes, deploys, and monitors the FL application at the central server, as shown in Figure~\ref{fig:fl_base}.  Every participating hospital collects X-rays of patients labeled by radiologists and trains a local ResNet model on that data. Hospitals periodically share their locally trained models with a central server. The central server then aggregates these shared models into one global model. After aggregation, the central server sends the updated global model to each hospital to incorporate in local training in the next round, as shown in Figure~\ref{fig:fl_base}.

The developer observes that multiple hospitals are reporting a high training loss from their preceding training rounds. One plausible reason is that one of the hospitals performed training on noisy data (mislabeled by inexperienced staff ~\cite{chen2020focus,li2019learning}) and continuously impacted the global model during aggregation. Thus, when the global model is shared back with the other hospitals, it influences their training.


\noindent{\textit{\textbf{Challenges of FL Debugging.}} After noticing an increase in training loss, the developer must investigate the root cause, as misdiagnosis from medical imaging can lead to ill treatment. To debug the FL application at this scale, the developer begins by manually inspecting various collected logs at the central server, including the global model weights, shared local models from hospitals, and the response and training time of each hospital. Due to patient privacy, the hospitals refrain from sharing their labeled training data, which is critical for correctly evaluating the quality of a model and thus essential for localizing the faulty round and model. Even if the developer finds the problematic round, she cannot isolate the hospital(s) responsible for affecting the global model without test data. One option is cross-validating each client's model by requesting that the other clients test the model on their local data. This is prohibited in practice, as it adds computational burden on clients (\eg edge devices) and can potentially cause data privacy violation. Lastly, statically inspecting hospitals' models does not provide any meaningful information. Without any debugging techniques at her disposal, she resorts to using guesswork to identify the hospital with noisy labels.

\noindent{\textit{\textbf{\tool's Contributions.}}} The developer decides to use \tool to investigate the root cause behind high training loss. When enabled, \tool allows a developer to set a {\em breakpoint} at any round or even in the first round to capture the end-to-end training logs. This breakpoint separately invokes a debugging session, a simulation of the original FL service, without stopping the live training.
In the debugging session, the developer uses \tool's  {\em step-back} and {\em step-next} constructs to move between rounds, inspecting the global and local models of hospitals.
Upon inspecting the training rounds, she finds the specific round, \eg~{\tt round 8}, where the performance starts to deteriorate. This round can be different from the {\em breakpoint} enabled round, as performance issues can manifest in earlier rounds but surface later.
During this inspection, \tool also reports the list of hospitals that participated in that round. 
Next, she invokes \tool's fault localization algorithm to precisely identify the hospital responsible for deteriorating the global model performance. After finding the hospital with noisy labels, the developer removes it from the problematic round (\ie~{\tt round 8}) and onwards. \tool's {\em fix and replay} starts retraining from {\tt round 8} to the current round and then replaces the impacted global model with the retrained global model and switches back to the original FL training. 
%


            

%% file: paper_sections/section_debugging-primitives.tex
\section{\tool's Debugging Constructs}
\label{debugging-primitives}

The goal of \tool is to facilitate an FL application developer in isolating a faulty client responsible for deteriorating the global model performance. Recent studies emphasize the need for debugging techniques in FL applications and the challenges associated with providing debugging support in FL frameworks~\cite{kairouz2021advances}. To this end, we must overcome the following {\em major challenges} in designing \tool. First, the privacy concerns of FL put restrictions on any client-side interference. Second, the unpredictable and ephemeral nature of clients in FL poses a threat to reproducibility, which is critical for debugging a live system. Third, the distributed nature of FL with hundreds of participating clients makes traditional breakpoint debugging ineffective. Pausing the entire FL application at this scale will be prohibitively expensive. Therefore, traditional debugging approaches, such as {\tt gdb}, are not suitable for the scale and architecture of FL systems.

In \tool, we address the above challenges and advance systematic FL application debugging. We enable realtime, interactive debugging on a simulation of the live FL application. To do so, \tool continuously collects and stores concise telemetry data from a live FL application. Whenever a debugging need arises, the developer can interact with \tool's debugging interface, which uses the telemetry data, to regenerate an FL application's state.

    \begin{figure}[t]
       \includegraphics[width=0.49\textwidth]{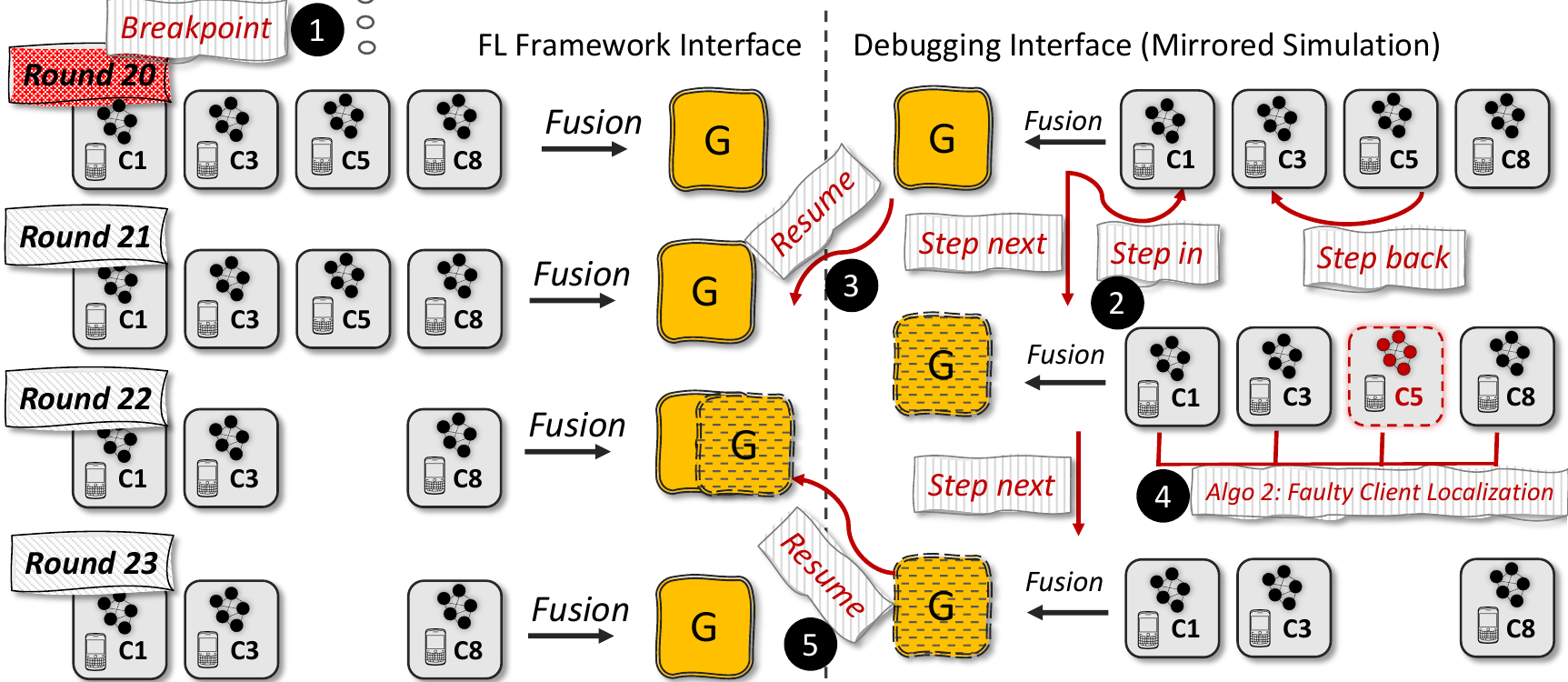}
        \caption{Using \tool, a developer can set a {\em breakpoint} at {\tt round 20}. When the FL application finishes {\tt round 20}, \tool launches a Debugging Interface, reflected on the right. {\em Step next} (\ding{203}) takes the developer to the next step (round or client). {\em Step-in} increases the granularity of computation, \eg round to client level. {\em Resume} (\ding{204}) rejoins the current execution status of the FL application if no intrusive actions are taken. At a given round, \tool can automatically localize the faulty client (\ding{205}) and then resume (\ding{206}) upon which the global model will be recomputed without the faulty client. This model will replace the corresponding round\textquotesingle s  model, and \tool will start retraining from that round, {\tt round 22}, in the FL interface.}
        \label{fig:breakpoint}
         \vspace{-4ex}
    \end{figure}
    
\subsection{Selective Telemetry}
\tool collects critical FL execution metrics to reproduce an FL application\textquotesingle s state for the developer to interact with it while investigating the root cause of a problem. Existing FL frameworks are carefully architected to refrain from revealing private data. As a result, most debugging data is private and cannot be investigated.

\tool's debugging approach is inspired by replay debugging. As with any other replay debugging approach, it is essential that \tool stores the necessary runtime metrics to reproduce an FL application\textquotesingle s state if requested by the developer. We design a highly selective FL event telemetry technique that records the concise execution data available at the central aggregator that is vital for generating any prior FL application state. 
\tool is different from traditional replay debugging as it only tracks the information needed to recreate an {\em observable} event and does not log the information unavailable to the developer in a live application. This design reduces the size of continuously growing telemetry data and minimizes the likelihood of information leakage. \tool mainly stores the information available after step 3 of Figure~\ref{fig:fl_base} which is clients' models, their reported metrics such as response time, training loss, validation loss, performance metric (\eg F1 score), hyperparameters (\eg learning rates, epochs, weight decay), and round ID. Note that the FL application, including client-side training, will continue uninterrupted in the background with \tool's telemetry module continuously collecting execution traces.

\subsection{Interactive Replay Debugging}
To start the interactive debugging process, a developer can invoke \tool's debugging constructs that let the developer leverage the telemetry data to investigate the root cause. Breakpoint debugging is the de-facto method of debugging a program. It pauses the program when the execution reaches it. At that point, a developer can inspect the values assigned to different variables, both local and global, and examine the method stack. Such debugging features are not applicable in FL. The traditional breakpoint will pause the distributed training, resulting in unnecessary idling at the client side. Additionally, since the state of a round is not saved,
it is currently impossible for the developer to inspect previous rounds. For instance, a developer may want to debug a latent issue that was introduced by a client five rounds ago but surfaced in the current round when the same client participated in training again.

We make the following observation about FL frameworks. An FL application only reveals  aggregator\textquotesingle s  events to a developer. In contrast, events on the client\textquotesingle s  side are entirely hidden from the developer except the ones relayed to the aggregator by the client. Building on this observation and the telemetry data captured by \tool, our insight is that instead of debugging a system in real-time, we can recreate its observable behavior in a simulated environment, giving an illusion of debugging an FL application in real-time. By doing so, inspections with \tool are side-effect free, \ie \tool will not interfere or interrupt the live FL application. Thus, eliminating the need to pause client-side training or halt FL aggregator execution.

\noindent{\textit{\textbf{Breakpoint.}}} To this end, \tool offers {\em breakpoint} that can help a developer inspect intermediate states of an FL application. \tool's breakpoint operates on computation units of {\em rounds}. 
Any abnormality in the client-reported metrics, such as training loss, validation loss, response time, and performance metrics (e.g., F1 score) can necessitate the use of breakpoints. \tool allows setting a breakpoint at any arbitrary round during live FL. A developer can also set a breakpoint from the start (\ie~{\tt round 0}) to capture end-to-end FL training traces or on a specific round (\eg~{\tt round 20} in Figure~\ref{fig:breakpoint}-\ding{202}) to inspect FL training at that round.
When the live FL application arrives at a breakpoint, \tool spawns a new debugging interface on the aggregator side, as shown in \ding{202} in Figure~\ref{fig:breakpoint}, while continuing the live FL training in the background.

%

\noindent{\textit{\textbf{Step in/Step out.}}} While at a breakpoint in a debugging session, a developer can use {\em step-in} and {\em step-out} actions to switch between different granularities of computational units. Traditionally, these two actions are used to go one-level deeper in the stack (\eg inside a function call) and move one level up in the stack (\eg outside the function call), respectively. Based on this convention, we define a round as a coarse-grained unit of computation that can be decomposed into a subset of clients participating in that round.
Suppose the current breakpoint is at {\tt round 20}. Step-in will take the developer to the clients-level granularity (\ding{203} in Figure~\ref{fig:breakpoint}) where trained models from clients are being aggregated, using a fusion algorithm (\eg FedAvg~\cite{mcmahan2017communication}). Step-out will take the developer back to the level of rounds, allowing them to inspect the global trained model at a higher level of abstraction and understand its performance across multiple rounds. Inspecting a state at client-level granularity entails evaluating the performance of a partially-aggregated global model. For example, in Figure~\ref{fig:breakpoint}, step-in at \ding{203} will take the execution between {\tt C1} and {\tt C3}, where the global model has yet to incorporate the local models of clients {\tt C5} and {\tt C8}.

\noindent{\textit{\textbf{Step Next/Step Back.}}} Similar to step-in/out, {\em step next} and {\em step back} help a developer transition from one state to another. For instance, if the breakpoint is at {\tt round 20}, step next will take the execution to {\tt round 21} in the debugging interface, showing information corresponding to that round only. Similarly, if the breakpoint is at client {\tt  C5}, step back will take the execution state to a partial global model after aggregating models from clients {\tt C1} and {\tt C3} only ({\tt Step back} in Figure~\ref{fig:breakpoint}).

\noindent{\textit{\textbf{Resume.}}} Unlike resume in {\tt gdb}, \tool's {\em resume} does not resume any paused execution. Instead, resume gives the illusion to the developer that execution is being continued from where it left off. \tool creates this environment by replaying the telemetry data that was captured while the FL application was being inspected using breakpoints, in case the developer does not find any faults in the round under inspection. Once the sequence of events in telemetry catches up with the live execution of the FL application, \tool switches to the FL interface and shuts down the debugging interface. This three-step process is nearly indistinguishable from an FL application with \tool disabled, giving the impression of debugging a real-time FL application interactively. {\em Resume} is also illustrated in Figure~\ref{fig:breakpoint} - \ding{204}.

\subsection{Fix and Replay}
\label{fix-and-rep}
When the developer successfully identifies a faulty client in any round, \tool offers {\em Fix and Replay} to allow a developer to roll back the training and provide a retrained global model (the one without a faulty client). We describe the technique to identify a faulty client in Section \ref{section:fault_localization}.  A faulty client may have a compound effect on the global model, as it may have begun to share its noisy model updates latently several rounds ago, which only later becomes noticeable. In such cases, it is important to rectify the impact of a faulty client\textquotesingle s inclusion in prior training rounds by removing its contributions. This requires retraining over multiple rounds, which is not possible as clients may not store the data used in training in the prior rounds. Figure~\ref{fig:breakpoint}-\ding{205} shows the removal of a faulty client ({\tt C5}) in {\tt round 21}. \tool recomputes the global model in the debugging interface and then replaces the actual global model in {\tt round 22} with the newly recomputed global model after fix and replay (Figure \ref{fig:breakpoint}-\ding{206}). By default, \tool forbids the faulty client from participating in the FL training. However, it is up to the developer to weigh the benefits of including the faulty client in future rounds.

%% file: paper_sections/section_fault_localization.tex
\section{Faulty Client Localization}
\label{section:fault_localization}

    \begin{figure}[t]
      \centering
           \includegraphics[width=0.50\textwidth]{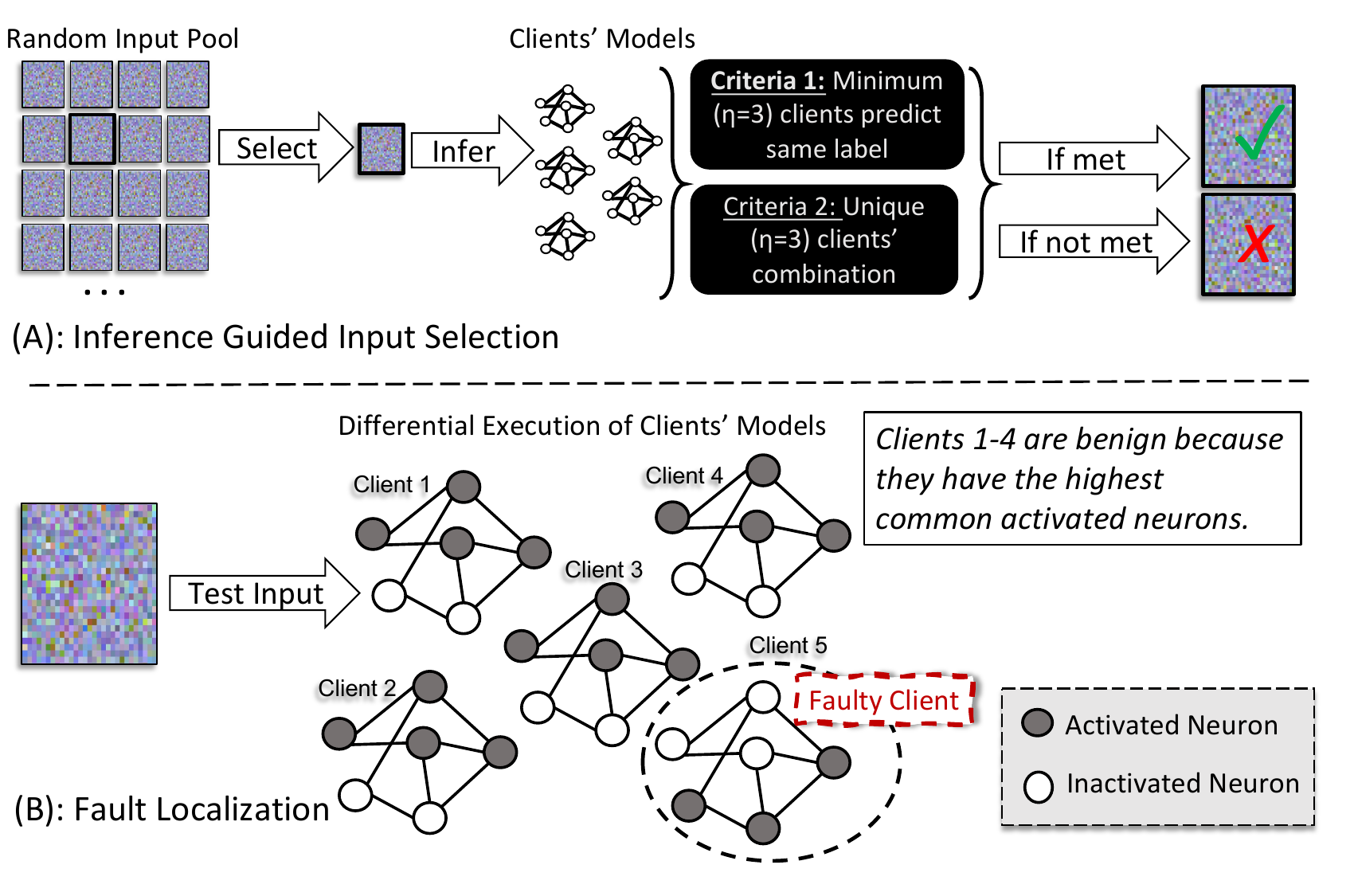}
            \vspace{-2ex}
            \caption{An overview of \tool's fault localization approach. It first selects a random input that invokes diverse model behavior (A). It then applies differential execution on clients' models to localize a faulty client (B).}
            \label{fig:fault_localization}
            \vspace{-2ex}
    \end{figure}

Faults in a client's model can arise due to measurement errors, human labeling errors, data poisoning, communication problems, or subjective biases of labellers.
For a high-quality global model, it is critical to correctly identify a faulty client and potentially restrict its participation. Manually identifying faulty clients is neither scalable nor effective due to a large number of participating clients in FL and their uninterpretable models. Furthermore, the model parameters (\ie~weights) do not provide any meaningful debugging information. To automate faulty client localization, we must define a feedback mechanism to guide our search for faulty clients efficiently. Automated debugging tools~\cite{zeller2002simplifying, le2011genprog} for regular software address this problem by relying on multiple test {\em inputs} and a test {\em oracle}. For example, unit tests can guide the search toward concise input leading to incorrect program output~\cite{zeller2002simplifying}. In FL, the inputs and oracle translate into diverse test data and the corresponding accurate labels, both of which are unavailable to the developer at the central server. 

\tool addresses the challenges of automated fault localization with a two-pronged approach. First, it generates a pool of random test inputs and applies a novel inference-guided test input selection to construct a suite of test inputs, as shown in Figure~\ref{fig:fault_localization}-A.
Since the test inputs are generated autonomously and are not accompanied by ground truth labels, metrics such as F1 score or accuracy cannot be used as oracle feedback to identify a faulty client.
Instead, \tool performs differential testing of clients' models to measure similarities and differences among models' behaviors on selected inputs (Figure~\ref{fig:fault_localization}-B). \tool fingerprints a neural network behavior on an input by profiling the internal neurons' contributions towards a model prediction. Subsequently, \tool accurately recognizes a client as faulty if its behavior deviates from the norm, which is the majority of the clients' behavior. Our insight is that a faulty client's model will show a noticeable difference in its internal neuron values compared to benign clients' models, based on the principle that faulty executions are intrinsically different from correct ones. The same principle is behind popular fault localization techniques, such as spectra-based fault localization~\cite{Jone2002sbfl} and delta debugging~\cite{zeller2002simplifying}.

\begin{algorithm}[t]
    {\footnotesize
    \SetKwInput{KwInput}{Input}                
    \SetKwInput{KwOutput}{Output}              
    \DontPrintSemicolon
    \KwInput{$shape$: dimension of the random input to be generated.}
    \KwInput{$\kappa$: number of inputs to be generated.}
    \KwInput{$\eta $: minimum number of clients for the same prediction.}
    \KwOutput{$\overline{X}$: {a list containing auto-generated test inputs.}} 
    
    \SetKwProg{Fn}{Function}{:}{}
    {
    \BlankLine
    \BlankLine
        $rand\_inputs = lazilyGenerateRandInputs(shape)$\;
        $\overline{X} = list( )$ \tcp{\scriptsize a list for inference guided test inputs}
        $seen\_clients\_sequences = list( )$

        {
                \While {$length(\overline{X}) < \kappa $}{
                    $ r\_input  = pop(rand\_inputs)$\;

                    $clients\_preds = getPredictions(clients, r\_input)$\;

                    \For{$label  \in class\_labels$}{
                        {\scriptsize $ clients\_seq =  samePredClients(clients\_preds, label)$}\;
                        \If{$clients\_seq \not \in  seen\_clients\_sequences$ \textbf{and} $ length(clients\_seq)$ $\geq$  $\eta $  }{
                            $seen\_sequences.append(clients\_seq)$\;
                            $\overline{X}.append(r\_input)$ \tcp{\scriptsize valid test input}
                            $\textbf{break}$
                        }

                    }

                    \If {$length(rand\_inputs) < 1 $}{
                        $rand\_inputs = lazilyGenerateRandInputs(shape)$\;
                    }
                }

                \KwRet $\overline{X}$\;
            }

    }
    \caption {Inference-Guided Test Input Selection}
    \label{algo:inputs}
    }
\end{algorithm}

\noindent{\textit{\textbf{Inference-Guided Test Input Selection.}}} As shown in Figure~\ref{fig:fault_localization}-A, \tool first lazily generates a pool of random test inputs using Kaiming Initialization~\cite{he2015delving}. For example, if the clients' models are trained on 32x32 images within the RGB scale, then \tool randomly creates a pool of synthetic inputs with the same size and format (\ie random images of size 32x32 in RGB scale). It then automatically selects only those inputs that lead to a consensus on predictions among a {\em unique} subset of clients. \tool selects up to $\kappa$ test inputs (default is $\kappa=10$) among the pool of 1000 random inputs. The goal is to minimize any overlapping behavior between clients while inferring unique class labels on selected test inputs. This is similar to achieving maximum code coverage in regular software with minimum tests. 
Algorithm~\ref{algo:inputs} selects a test input (line 5) if at least ($\eta \geq$ 5) clients predict the same label and that subset of clients has not been seen in a previously selected input (lines 6-11). 
On the next random input, if the previously observed subset of clients (\ie $clients\_seq \in seen\_clients\_sequences$) predict the same class label, we discard this input. If a unique combination of clients predicts an unseen label, we include the input in the test suite. This process is repeated until we collect a user-defined, $\kappa$, number of test inputs.

\noindent{\textit{\textbf{Differential Execution of Clients Models.}}} In the absence of correct labels of generated test inputs, \tool adapts differential testing to find behavioral differences and similarities among clients' models, as shown in Figure~\ref{fig:fault_localization}-B. \tool profiles the contributions of individual neurons during model inference on an input and uses these neurons activations to identify models with common behavior. Note that clients' models in FL are comparable due to having a similar architecture. Algorithm~\ref{algo:detection} describes the faulty client localization process. For a selected test input, \tool exhaustively iterates all possible combinations of potentially non-faulty clients (\ie $n \choose 1$ combinations). For each combination, Algorithm~\ref{algo:detection} performs model inference on the test input and captures its neuron profiles. \tool aims to find one combination of clients that has the highest overlap in behavior, representing the true $n-1$ benign clients and consequently isolating the precise faulty client. This is a lightweight process due to the negligible model inference time and the iterations' linear time ($O(n)$) complexity.

Our insight is that among all possible combinations of clients, only one represents true benign clients' subset. The remaining combinations contain the faulty client with abnormal neuron activations, reducing the model behavior overlap within that set. In summary, at a given ill-performing round in FL, \tool takes in all participating clients' models as the only input. It automatically generates test inputs and employs differential testing on clients' models to monitor abnormal behavior to precisely identify a faulty client.

\begin{algorithm}[t]

    {\footnotesize
     \SetKwInput{KwInput}{Input}                
     \SetKwInput{KwOutput}{Output}              
    \DontPrintSemicolon
    \KwInput{$clients$: a list of clients participated in the given FL round.}
    \KwInput{$x$: a random input belongs to $\overline{X}$.}
    \KwInput{$na\_t$: a threshold to profile neuron activations.}
    \KwOutput{$faulty\_client$: {\scriptsize the faulty client who has abnormal behavior.}}
    
    \SetKwProg{Fn}{Function}{:}{}
    {
        {
        \BlankLine
        \BlankLine
                $all\_clients\_combinations = nChooseK(clients,1$)\;
                $benign\_clients = set()$\;
                $max\_common\_activations = -1$\;
                
                \For{$t\_clients \in all\_clients\_combinations$}{
                    $neuron\_ids = ActivatedNeurons(t\_clients, x, na\_t)$\; 
                    $t\_clients\_common\_neurons = intersection(neuron\_ids)$\;
                    $temp\_n = length(t\_clients\_common\_neurons)$\;
                    \If{$temp\_n > max\_common\_activations$  }{
                        $max\_common\_activations = temp\_n$\;
                        $benign\_clients = t\_clients$\;
                     }
                }
            }
                        
            $faulty\_client = clients - benign\_clients$\; 
            \KwRet $faulty\_client$\;
        }
        
      }
    \caption {Faulty Client Localization using Differential Testing}
    \label{algo:detection}
    \end{algorithm}

%% file: paper_sections/section_evaluation.tex
\section{Evaluation}
\label{evaluation}

We evaluate \tool on (1) runtime performance overhead, (2) debugging time, (3) fault localizability, and (4) scalability. Our evaluation aims to answer the following research questions:

    \begin{itemize}
        \item {\bf RQ1.} What impact does \tool have on the baseline FL framework's performance? 
        \item {\bf RQ2.} How accurate is \tool in identifying a faulty client? 
        \item {\bf RQ3.} Can \tool identify multiple faulty clients?
        \item {\bf RQ4.} Can \tool scale to large number of clients and find a faulty client efficiently?

    \end{itemize}
        
\noindent{\textit{\textbf{Datasets, Models, and FL Framework.}}} We evaluate \tool on CIFAR-10 and FEMNIST datasets. Both are considered as gold standard to evaluate FL experimental settings~\cite{collins2021exploiting, li2022federated}.
FEMNIST is a modified version of MNIST presented in the FL LEAF Benchmark~\cite{caldas2018leaf} and the Non-IID Bench~\cite{li2022federated}. The FEMNIST dataset includes more than 340K training and 40K testing grayscale images, each with a resolution of 28x28 pixels, representing ten distinct class labels. CIFAR-10 contains 50K training 32x32 RGB images that span ten different classes and 10K instances for testing. 
We adopt popular CNN models, namely ResNet~\cite{he2015deep}, VGG~\cite{simonyan2014very}, and DenseNet~\cite{huang2017densely} architectures. 
We set the learning rate between 0.0001 and 0.001, the number of  epochs between 10 and 25, the batch size from 512 to 2048, and the weight decay to 0.0001. We realize \tool's design in the IBMFL library~\cite{ibmfl2020ibm} due to its ease of use, open documentation, and publicly available codebase. These techniques should be equally applicable to other FL frameworks. 

\noindent{\textit{\textbf{Evaluation Environment Specifications.}}} We run our experiments on an AMD 16-core processor with 128 GB RAM and an NVIDIA Tesla T4 GPU. To measure the performance of \tool in terms of runtime and debugging overhead, we simulate IBMFL framework deployment on a MacBook Pro with a Quad-core Intel Core i5 processor and 16 GB RAM.

\noindent{\textit{\textbf{Federated Learning Experimental Settings.}}} Prior FL literature~\cite{caldas2018leaf, li2022federated} establishes two data distribution strategies among FL clients: IID (independent and identically distributed data) and Non-IID (non-independent and identically distributed data). For Non-IID, we use the quantity base imbalance~\cite{li2022federated} where clients have an unequal quantity of data, and the class distribution is random. In IID, the clients receive the same quantity of data. None of the clients share the same data points in both settings. We simulate FL with a varying number of clients, ranging from 10 to 400 clients, in each FL training round. In practice, even with millions of clients, only a subset (in the order of hundreds) is selected in a round. Therefore, our experiment settings are representative of real-world FL deployments~\cite{li2022federated, mcmahan2017communication, wang2020tackling, avdiukhin2021federated, bonawitz2019towards, 45648}.


\noindent{\textit{\textbf{Fault Injection.}}} Since there is no existing FL benchmark with faulty clients, \tool adopts a standard noisy labels approach from prior machine learning literature to inject a faulty client in our experiments~\cite{yao2018deep, jiang2020beyond, li2021learning, frenay2013classification, hendrycks2018using}. Similar to prior work~\cite{ma2018dimensionality, ghosh2017robust, li2020dividemix}, we arbitrarily add noise by changing training data labels (\eg changing label \say{bird} to \say{cat}). When such a client's model is merged with the global model, the global model's performance (\eg accuracy) deteriorates. We define different strengths of a fault with a {\em noise rate} that controls the number of labels modified in a faulty client. Noise rate is defined as a ratio between changed labels and original labels ({\em changed-labels/original-labels}).

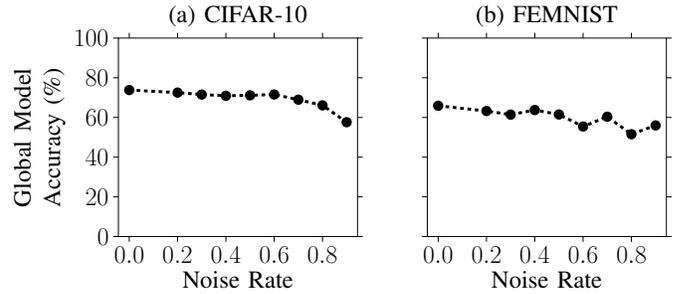
\begin{figure}[t]
    \vspace{-2ex}
    \centering
    \resizebox{\linewidth}{!}
    {\input{graphs/tikz_global_acc}}
    \caption{Global model (ResNet-34) prediction accuracy in the presence of a faulty client with different noise rates. Lower noise rates hardly degrade global model performance.}
    \label{fig:flipped_labels_global_model_performance}
    \vspace{-2ex}
\end{figure}

Figure~\ref{fig:flipped_labels_global_model_performance} shows the impact of different noise rates on the global model's accuracy, with one faulty client and nine benign clients. 
Low noise rates, ranging from 0.2 to 0.7, barely affect the global model performance. With a 0.7 noise rate, the accuracy is lowered by 4.8\% and 5.5\% in CIFAR-10 and FEMNIST, respectively. A noise rate of 0.9 incurs a 16.2\% and 9.9\% reduction in the global model accuracy in both settings. Thus, to have a measurable impact on the global model's performance, we select a noise rate of one for a faulty client. Still, we perform sensitivity analysis in Section~\ref{subsec:localize-faulty-clients} (Figure~\ref{fig:faults-2-9}) by measuring the impact of varying noise rates on \tool's fault localizability.

\noindent{\textit{\textbf{Neuron Activation Threshold.}}} We adopt the method from Harel-Canada et al.~\cite{harel2020neuron} to profile neuron activations. We empirically find 0.003 as the optimal value for the default activation threshold (see Section~\ref{subsec:neuron-activation}). A neuron is considered active when its value crosses this threshold.

\noindent{\textit{\textbf{Faulty Client Localization Accuracy.}}} We calculate faulty client localization accuracy as the ratio between (a) the number of test inputs on which faulty clients are correctly identified and (b) the total number of test inputs. For instance, if \tool identifies the correct set of faulty clients on four out of ten test inputs generated by Algorithm~\ref{algo:inputs}, we report 40\% fault localization accuracy.

\subsection{FedDebug's Performance}
Capturing telemetry data in realtime may slow down the performance of an FL application's aggregator. In this subsection, we present the evaluation results of \tool's runtime overhead and the fault localization time. These experiment settings employ ResNet-18 with CIFAR-10.

\noindent{\textit{\textbf{Runtime Overhead (RQ1).}}} To evaluate the impact on the FL application's performance, we measure the slowdown in the running time that \tool incurs. We compare the cumulative processing time of the vanilla IBMFL's aggregator (baseline) against that of the \tool-enabled aggregator on a variety of client combinations, ranging from 5 clients to 100 clients. The aggregation time varies with the model's architecture and the number of clients participating in a round, but it is completely independent of the models' quality. Therefore, we create up to 100 pre-trained ResNet-18 models and perform the aggregation.

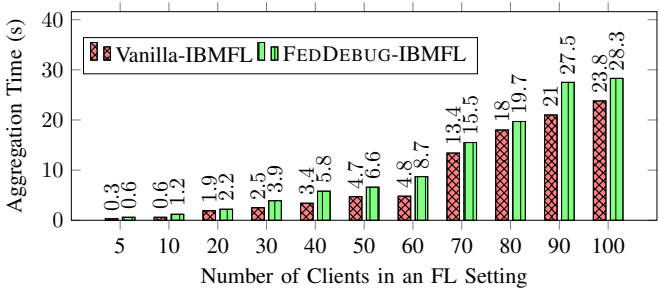
\begin{figure}[ht]
 \centering
    \resizebox{\linewidth}{!}
    {\input{graphs/runtime-overhead}}	
    \caption{\tool's runtime overhead as a comparison between vanilla FL framework's aggregation time with \tool enabled FL aggregation. }
	\label{plot:runtime-overhead} 
    \vspace{-2ex}
\end{figure}

Figure~\ref{plot:runtime-overhead} compares the baseline's aggregation time with the \tool enabled aggregation time. The X-axis represents the number of clients ranging from 5 to 100 clients, and the Y-axis represents the average time across two FL rounds. For instance, with 30 clients, \tool takes 3.9 seconds compared to the 2.5 seconds for the baseline to aggregate 30 trained models into a global model. Overall, \tool takes approximately 48\% additional aggregation time across all experiments. However, in an end-to-end round, the training phase on the clients' end occupies the majority (up to 97.8\% in our experiments) of the round's time. Compared to the training time of a round, the aggregation time is almost negligible, as low as 1.2\% in our experiments.

    \begin{tcolorbox}[left=0mm, right=0mm, top=0mm, bottom=0mm]
    \textbf{Summary:} Considering the training and aggregation time of each FL round, \tool's runtime overhead is a very small fraction, 1.2\%, of the training time. Hence, capturing telemetry data for replay debugging does not impede the FL application's runtime performance.
    \end{tcolorbox}

\noindent{\textit{\textbf{Debugging Time (RQ1).}}} To assess the localizability of \tool, we design experiments to measure \tool's \emph{debugging time}, the time it takes to localize a faulty client. We then compare this time with the training time of that round. Since there is no comparable approach to localize a faulty client, we use training time as a baseline to provide a meaningful scale for the cost of debugging.

\begin{figure}[t]
        \centering
        
        \resizebox{\linewidth}{!}
        {\input{graphs/tikz_Algo1_Algo2_Train.tex}}
        \caption{\tool's debugging time contains input generation time and faulty client detection time and is compared against a round's training time.}
        \label{plot:debugging-overhead}
    \end{figure}
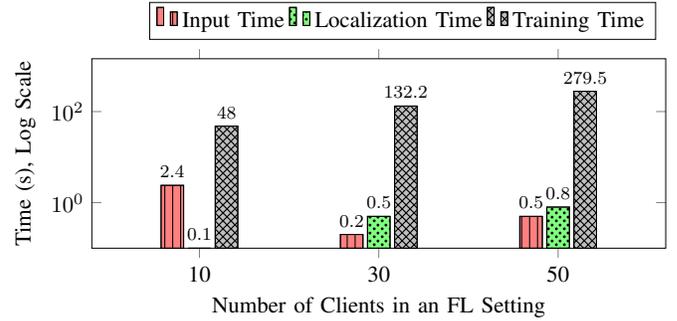

Figure~\ref{plot:debugging-overhead} shows the results of these experiments. The X-axis represents the number of clients, and the Y-axis shows the debugging time in seconds on a logarithmic scale. For 30 clients, \tool's input generation and selection takes 0.2 seconds to find high-quality test input, and its fault localization takes approximately 0.5 seconds to localize a faulty client. In a ten clients setting, input selection takes longer due to  constraint  $\eta = 4 $ for criteria 1 in Figure~\ref{fig:fault_localization}. $\eta = 4$ means that at least four previously unseen clients should predict the same label on newly selected test input.

Overall, our results show an increasing debugging time when the number of clients increases, which is expected as increasing the number of clients increases the search space. Note that the debugging time is still in the order of seconds, even for 50 clients. This is because 1) for {\tt n} clients, the search space has at most {\tt n} possible combinations of potentially benign {\tt n-1} clients, representing linear complexity, and 2) on a given input, \tool only profiles neuron activations once while iterating over the {\tt n} combinations.

\begin{tcolorbox}[left=0mm, right=0mm, top=0mm, bottom=0mm]
\textbf{Summary:} On average, \tool can efficiently identify a faulty client in 2.1\% of the total training time of a round.
\end{tcolorbox}

\subsection{Localization of Faulty Client(s)}
\label{subsec:localize-faulty-clients}

    \begin{table}[t]
    {    
     \caption{\tool's debugging time and accuracy when localizing a faulty client in 36 different FL settings with 100 test inputs.}
    
        \centering
        \scalebox{.85}{
\input{tables/table_flip_10}  
        }
        
        \label{table:flip_all_labels}

    } 
    \end{table}

To answer RQ2, we measure how accurate \tool is in localizing a faulty client. We inject a faulty client that is representative of a real-world scenario and can cause a measurable change in the global model's performance. By varying the number of clients, datasets, models, and data distributions (IID and Non-IID), we create 36 different FL configurations for \tool's evaluation.

Column 4 and 5 of Table~\ref{table:flip_all_labels} show the accuracy of \tool in the IID and Non-IID settings, respectively. We repeat each experiment on 100 generated test inputs and take the average of each metric to generalize the results. \tool correctly identifies a faulty client with 100\% accuracy in both IID and Non-IID settings.

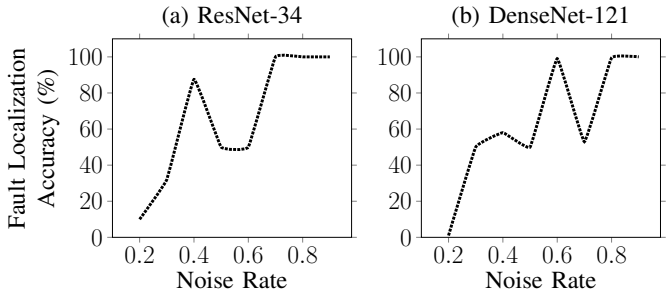
\begin{figure}[t]
        \centering
 \resizebox{\linewidth}{!}
        {\input{graphs/tikz_noise_rate_fault_detection.tex}}
        \caption{\tool localization performance when a faulty client has varying fault strength (\ie low noise rate). }
        \label{fig:faults-2-9}
        \vspace{-2.5ex}
    \end{figure}

\noindent{\textit{\textbf{Varying Noise Rate.}}} Figure~\ref{fig:flipped_labels_global_model_performance} shows the impact of different noise rates on the global model prediction accuracy. We observe that a faulty client has a measurable impact on the global model with a noise rate of $> 0.8$.
    The global model's accuracy merely drops from 73.8\% to 71.1\% when the faulty client has a 0.6 noise rate, and drops to 57\% when the noise rate is close to one. 
    \tool localizes faulty client(s) with low noise rates, showing its robustness. Figure~\ref{fig:faults-2-9} shows the evaluations on varying noise rates in 10 clients FL settings with ResNet and DenseNet architectures. The X-axis shows the faulty client's noise rate, and the Y-axis represents the average fault localization accuracy on the CIFAR-10 and FEMNIST datasets. The results, as seen in Figure~\ref{fig:faults-2-9}, indicate that \tool has the capability to identify low noise faults---it successfully localizes a faulty client with 0.4 noise rate with approximately 58\% and 87.5\% accuracy in DenseNet and ResNet settings, respectively.

    \begin{tcolorbox}[left=0mm, right=0mm, top=0mm, bottom=0mm]
    \textbf{Summary:} \tool achieves 100\% fault localization accuracy on average on a total of 3600 test inputs when the faulty client significantly deteriorates the global model performance in both IID and Non-IID settings. It also accurately localizes a faulty client with low noise rates. 
    \end{tcolorbox}

\noindent{\textit{\textbf{Detecting Multiple Faulty Clients (RQ3).}}}
    We evaluate \tool's ability to identify multiple faulty clients. To this end, we inject up to seven faulty clients in the following experiment settings. We train ResNet-50 and DenseNet-121 on the CIFAR-10 and FEMNIST datasets in 30 and 50 clients FL settings. Each setting is evaluated on 10 test inputs. By default, \tool's fault localization technique finds a single faulty client. We apply \tool in an iterative manner to find multiple faulty clients by removing one faulty client on each iteration, similar to traditional bug repair process, where one bug is fixed first before the next one is investigated.

    Table~\ref{table:corrupt} presents the results of finding multiple faulty clients in 32 FL configurations. For instance, when 7 out of 30 clients are faulty and the model is ResNet-50, \tool finds all seven faulty clients with 100\% accuracy on CIFAR-10 and 97.1\% accuracy on FEMNIST. Compared to ResNet, \tool performs relatively better with DenseNet. This behavior is expected because, compared to ResNet, DenseNet learns better features due to dense concatenation among its layers, resulting in better performance~\cite{zhang2021resnet}. 
    Thus, \tool performs well in localizing multiple faults with DenseNet with an average accuracy of 99.7\% on both datasets compared to ResNet's 80.8\%.

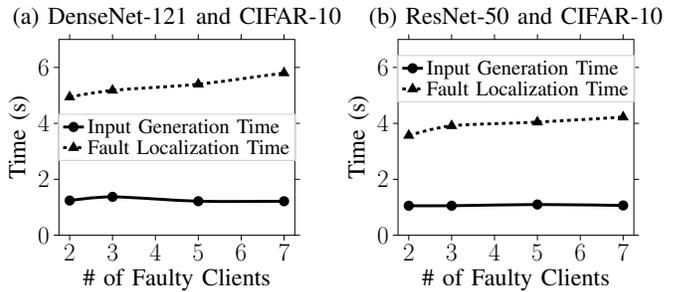
\begin{figure}[t]
        \centering
      \resizebox{\linewidth}{!}
        {\input{graphs/tikz_multi_fault_overhead.tex}}
        \caption{\tool finds multiple faulty clients in a linear time. Total clients are 50 in each graph.}
        \label{fig:multi_faulty_cleint_overhead}
        \vspace{-3ex}
    \end{figure}

    Table~\ref{table:corrupt} also reveals that, generally, \tool's localization performance is positively correlated to the number of training data points per client. Large, high-quality training data promotes better feature learning among neurons and, thus, yields better performance. Since the number of data points in FEMNIST (340K) is large compared to CIFAR-10 (40K), clients in the FEMNIST settings have significantly larger training data than clients in the CIFAR-10  settings. As a result, \tool average localization accuracy is 78.5\% in the ResNet-CIFAR experiment, while it has 83.1\% localization accuracy in the ResNet-FEMNIST experiment. \tool finds multiple faults with linear time complexity, as shown in Figure~\ref{fig:multi_faulty_cleint_overhead} with 50 clients. The input generation time is almost constant, as the number of clients is fixed. However, the localization time increases as we increase the number of faults from 2 to 7. For instance, it localizes two faulty clients in 3.6 seconds and five faulty clients in 4 seconds.
    

     \begin{table}[t]
             \caption {\tool's fault localization in 32 FL configurations with multiple faulty clients, ranging from two to seven.} 
    {    
        \centering
        \scalebox{.99}{
          \input{tables/table_corrupt}

        }        
        \label{table:corrupt}
    } 
    \end{table}

\begin{figure}[t]
        \centering
        \resizebox{\linewidth}{!}
        {\input{graphs/tikz_scalability_400.tex}}
        \caption{\tool retains scalability on a large number of clients.}
        \label{plot:scalability}
        \vspace{-2.5ex}
    \end{figure}
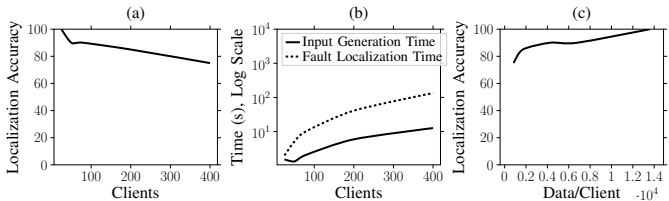

\noindent{\textit{\textbf{Scalability (RQ4).}}} Our findings also show that \tool scales to larger datasets and an increasing number of clients in FL. Figure~\ref{plot:scalability} summarizes the impact on \tool's ability to identify a faulty client when the number of clients changes from 25 to 400 and the training data size per client changes. We perform this experiment with two faulty clients in the FEMNIST-DenseNet configuration. Figure~\ref{plot:scalability}-(a) verifies that \tool's fault localization accuracy only reduces to 75\% even when the number of clients increases to 400. \tool's debugging time increases linearly as the number of clients increases, consistent with the scale-up properties of general distributed systems, as shown in Figure~\ref{plot:scalability}-(b). When the number of clients increases, less data is used to train a client's model, which may reduce the accuracy of clients' models. Figure~\ref{plot:scalability}-(c) also shows that \tool's fault localizability also increases when the number of data points per client increases, and it is also robust against low performing client models. For instance, when the number of data points increases from 850 to 1700, \tool's localization accuracy also changes from 75\% to 85\%, respectively.

    \begin{tcolorbox}[left=0mm, right=0mm, top=0mm, bottom=0mm]
        \textbf{Summary:} Our experiment results provide concrete evidence that \tool preserves scalability properties both in terms of time overhead and in the presence of multiple faults. It successfully identifies multiple faulty clients in 32 different FL configurations with an average accuracy of 90.3\%.
    \end{tcolorbox}

\subsection{Neuron Activation Threshold}
\label{subsec:neuron-activation}

There is no standard threshold of neuron activations~\cite{pei2017deepxplore} and prior work uses experiential value for different use cases~\cite{harel2020neuron}. We evaluate the impact of different activation thresholds on \tool's faulty client localizability. We take 30 clients including five faulty clients, and train ResNet-50 and DenseNet-121 on both the CIFAR-10 and FEMNIST datasets. We repeat each experiment on 10 different inputs generated by Algorithm~\ref{algo:inputs}. 

\begin{figure}[t]
        \centering
        \resizebox{\linewidth}{!}
        {\input{graphs/tikz_nc_thresholds.tex}}
        \caption{\tool performance at neuron activation threshold on 30 clients, including five faulty clients.} 
        \label{fig:thresholds}
        \vspace{-2.5ex}
    \end{figure}
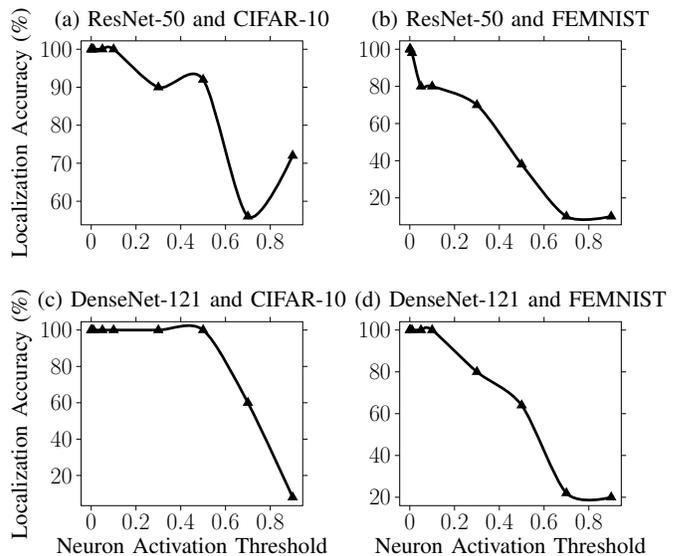

Figure \ref{fig:thresholds} shows the result of these experiments. The X-axis represents the neuron activation thresholds, ranging from 0 to 0.9. The Y-axis shows the \tool's localization accuracy in a given experiment setting. For instance, at the 0.003 threshold, the average localization accuracy across four settings is 100\%. On the other hand, at 0.5 threshold, the average accuracy decreases significantly to 73.5\% across these configurations. Specifically, for DenseNet-121 and FEMNIST experiment in Figure~\ref{fig:thresholds}-(d), the localization drops to 64\% at the 0.5 neuron activation threshold. We observe that \tool performs better at lower thresholds ($<$ 0.01) across different models and datasets. This behavior is expected because lower thresholds increase the sensitivity of \tool's localization approach. It starts monitoring most of the neurons' compared to a higher threshold, where \tool profiles only a few neurons crossing the threshold.

%% file: graphs/tikz_global_acc.tex
\begin{tikzpicture}

\definecolor{darkgray176}{RGB}{176,176,176}

\begin{groupplot}[group style={group size=2 by 1, horizontal sep=2cm}]
\nextgroupplot[
tick align=outside,
tick pos=both,
title={(a) CIFAR-10},
x grid style={darkgray176},
xlabel={Noise Rate},
title style={font=\huge},
xmin=-0.045, xmax=0.945,
xtick style={color=black},
xtick={-0.2,0,0.2,0.4,0.6,0.8,1},
xticklabels={
  \(\displaystyle {\ensuremath{-}0.2}\),
  \(\displaystyle {0.0}\),
  \(\displaystyle {0.2}\),
  \(\displaystyle {0.4}\),
  \(\displaystyle {0.6}\),
  \(\displaystyle {0.8}\),
  \(\displaystyle {1.0}\)
},
y grid style={darkgray176},
ylabel={Global Model \\Accuracy (\%)},
ylabel style={align=center},
ymin=0, ymax=100,
ytick style={color=black},
ytick={0,20,40,60,80,100},
label style={font=\huge},
tick label style={font=\huge},
yticklabels={
  \(\displaystyle {0}\),
  \(\displaystyle {20}\),
  \(\displaystyle {40}\),
  \(\displaystyle {60}\),
  \(\displaystyle {80}\),
  \(\displaystyle {100}\)
}
]
\addplot [line width=2.4pt, black, dashed, mark=*, mark size=3, mark options={solid}]
table {%
0 73.820001
0.2 72.530001
0.3 71.55
0.4 70.94
0.5 71.160001
0.6 71.579999
0.7 68.949997
0.8 66.100001
0.9 57.620001
};

\nextgroupplot[
scaled y ticks=manual:{}{\pgfmathparse{#1}},
tick align=outside,
label style={font=\huge},
tick label style={font=\huge},
tick pos=both,
title={(b) FEMNIST},
title style={font=\huge},
x grid style={darkgray176},
xlabel={Noise Rate},
xmin=-0.045, xmax=0.945,
xtick style={color=black},
xtick={-0.2,0,0.2,0.4,0.6,0.8,1},
xticklabels={
  \(\displaystyle {\ensuremath{-}0.2}\),
  \(\displaystyle {0.0}\),
  \(\displaystyle {0.2}\),
  \(\displaystyle {0.4}\),
  \(\displaystyle {0.6}\),
  \(\displaystyle {0.8}\),
  \(\displaystyle {1.0}\)
},
y grid style={darkgray176},
ymin=0, ymax=100,
ytick style={color=black},
yticklabels={}
]
\addplot [line width=2.4pt, black, dashed, mark=*, mark size=3, mark options={solid}]
table {%
0 65.894395
0.2 63.249415
0.3 61.432213
0.4 63.746572
0.5 61.508131
0.6 55.441809
0.7 60.374218
0.8 51.555151
0.9 55.995297
};
\end{groupplot}

\end{tikzpicture}

%% file: graphs/runtime-overhead.tex
\begin{tikzpicture}
\begin{axis}[
symbolic x coords={5, 10,20,30, 40, 50,60,70,80,90,100},
	xtick=data,
    width=0.6\textwidth,
    height=0.27\textwidth,
    bar width = 0.2cm,
	ylabel={Aggregation Time (s)},
	xlabel= {Number of Clients in an FL Setting},
	 legend pos=north west,
     legend columns=3,
    legend style={at={(0.02,0.70)}, anchor=south west},
	ybar ,
   label style={font=\Large},
   ymin=0,
   ymax=40,
	  ylabel near ticks,
        point meta=rawy,
           xlabel near ticks,
            every node near coord/.append style={rotate=90, anchor= west},
           nodes near coords,
          enlarge x limits=0.10, 
          enlarge y limits={value=.04,upper},
]

\addplot[fill={white!50!red}, postaction={pattern=crosshatch}] table [x=Parties, y=BaseAgg, col sep=comma] {overhead.csv};
\addplot[fill={white!50!green}, , postaction={pattern=vertical lines}] table [x=Parties, y=FedDebugAgg, col sep=comma] {overhead.csv};
\legend{Vanilla-IBMFL, \tool-IBMFL}
\end{axis}
\end{tikzpicture}

%% file: graphs/tikz_Algo1_Algo2_Train.tex
\begin{tikzpicture}
	\begin{axis}[
			symbolic x coords={10, 30, 50},
			xtick=data,
			 width=0.58\textwidth,
			 height=0.25\textwidth,
			ylabel={Time (s), Log Scale},
			xlabel= {Number of Clients in an FL Setting},
			legend pos=north west,
			legend columns=3,
			legend style={at={(.1,1.1)}, anchor=south west},
			ybar ,
			label style={font=\Large},
			ymax=1000,
			ylabel near ticks,
			ymode = log,
			log origin=infty,
			point meta=rawy,
			xlabel near ticks,
			every node near coord/.append style={rotate=0,font=\footnotesize, anchor= south},
			nodes near coords,
			enlarge x limits=0.30, 
			enlarge y limits={value=.04,upper},
		]
		
		\addplot[fill={white!50!red}, postaction={pattern=vertical lines} ] table [x=Clients, y=Algo. 1 Time, col sep=comma] {algo1_2_train.csv};
		\addplot[fill={white!50!green}, postaction={pattern=crosshatch dots}] table [x=Clients, y=Algo. 2 Time, col sep=comma] {algo1_2_train.csv};
		
		\addplot[fill={white!50!gray}, postaction={pattern=crosshatch} ] table [x=Clients, y=Training Time, col sep=comma] {algo1_2_train.csv};
		
		\legend{Input Time, Localization Time, Training Time}
	\end{axis}
\end{tikzpicture}

%% file: tables/table_flip_10.tex
\begin{tabular}{p{0.7cm}p{1.15cm} l
%
>{\raggedleft\arraybackslash}p{1cm}
>{\raggedleft\arraybackslash}p{1cm}
>{\raggedleft\arraybackslash}p{1cm}
>{\raggedleft\arraybackslash}p{1.2cm}}
\toprule
 Clients &   Dataset &  Architecture &  Accuracy \% (IID) &  Accuracy \% (Non-IID) &  Avg. Input Time (s) &  Avg. Localization Time (s) \\
\midrule
      10 &  CIFAR10 &  DenseNet-121 &             100 &              100 &                 2.41 &                     0.44 \\
      10 &  CIFAR10 &     ResNet-50 &             100 &              100 &                 2.40 &                     0.22 \\
      10 &  CIFAR10 &        VGG-16 &             100 &              100 &                 2.40 &                     0.21 \\
      30 &  CIFAR10 &  DenseNet-121 &             100 &              100 &                 2.42 &                     1.29 \\
      30 &  CIFAR10 &     ResNet-50 &             100 &              100 &                 1.18 &                     0.70 \\
      30 &  CIFAR10 &        VGG-16 &             100 &              100 &                 2.41 &                     0.47 \\
      50 &  CIFAR10 &  DenseNet-121 &             100 &              100 &                 2.42 &                     3.26 \\
      50 &  CIFAR10 &     ResNet-50 &             100 &              100 &                 1.37 &                     1.24 \\
      50 &  CIFAR10 &        VGG-16 &             100 &              100 &                 2.43 &                     0.91 \\
      10 &   FEMNIST &  DenseNet-121 &             100 &              100 &                 2.40 &                     0.47 \\
      10 &   FEMNIST &     ResNet-50 &             100 &              100 &                 2.40 &                     0.25 \\
      10 &   FEMNIST &        VGG-16 &             100 &              100 &                 2.40 &                     0.18 \\
      30 &   FEMNIST &  DenseNet-121 &             100 &              100 &                 2.41 &                     1.37 \\
      30 &   FEMNIST &     ResNet-50 &             100 &              100 &                 0.91 &                     0.68 \\
      30 &   FEMNIST &        VGG-16 &             100 &              100 &                 2.41 &                     0.55 \\
      50 &   FEMNIST &  DenseNet-121 &             100 &              100 &                 2.24 &                     2.44 \\
      50 &   FEMNIST &     ResNet-50 &             100 &              100 &                 1.42 &                     1.24 \\
      50 &   FEMNIST &        VGG-16 &             100 &              100 &                 2.40 &                     1.25 \\
\bottomrule
\end{tabular}

%% file: graphs/tikz_noise_rate_fault_detection.tex
\begin{tikzpicture}
  \begin{groupplot}[
      group style={group size=2 by 1, horizontal sep= 2cm},
      xmin=0,
      ymin=0,
      xlabel=Noise Rate,
      tick label style={font=\huge},
      label style={font=\huge},
      title style={font=\huge},
      xmin=0.1,
    tick align=outside, 
      tick pos=both,
    ]
    \nextgroupplot[title=(a) ResNet-34, ylabel= Fault Localization \\Accuracy (\%), ylabel style={align=center}]
    \addplot [line width=2.4pt, densely dotted, black, smooth,tension=0.15]  table [x=noise_rate, y=yResnet34, col sep=comma] {noise_rate.csv};
    \nextgroupplot [title=(b) DenseNet-121]
    \addplot [line width=2.4pt, densely dotted, black, smooth,tension=0.15]  table [x=noise_rate, y=yDensenet121, col sep=comma] {noise_rate.csv};
  \end{groupplot}
\end{tikzpicture}


%% file: tables/table_corrupt.tex
\begin{tabular}{p{1cm}p{1cm}l
>{\raggedleft\arraybackslash}p{1.6cm}
>{\raggedleft\arraybackslash}p{1.6cm}}

\toprule
 Faulty Clients &  Total Clients &  Architecture &  Accuracy \% (CIFAR-10) &  Accuracy \% (FEMNIST) \\
\midrule
                    2 &             30 &     ResNet-50 &               100 &              100 \\
                    3 &             30 &     ResNet-50 &               100 &              100 \\
                    5 &             30 &     ResNet-50 &               100 &               98 \\
                    7 &             30 &     ResNet-50 &               100 &               97.1 \\
                    2 &             30 &  DenseNet-121 &               100 &              100 \\
                    3 &             30 &  DenseNet-121 &               100 &              100 \\
                    5 &             30 &  DenseNet-121 &               100 &              100 \\
                    7 &             30 &  DenseNet-121 &               100 &              100 \\
                    2 &             50 &     ResNet-50 &                50 &               80 \\
                    3 &             50 &     ResNet-50 &                66.7 &               66.7 \\
                    5 &             50 &     ResNet-50 &                54 &               60 \\
                    7 &             50 &     ResNet-50 &                57.1 &               62.9 \\
                    2 &             50 &  DenseNet-121 &               100 &              100 \\
                    3 &             50 &  DenseNet-121 &               100 &              100 \\
                    5 &             50 &  DenseNet-121 &               100 &              100 \\
                    7 &             50 &  DenseNet-121 &               100 &               95.7 \\
\bottomrule
\end{tabular}

%% file: graphs/tikz_multi_fault_overhead.tex
\begin{tikzpicture}

    \definecolor{darkgray176}{RGB}{176,176,176}
    \definecolor{lightgray204}{RGB}{204,204,204}

    \begin{groupplot}[group style={group size=2 by 1, horizontal sep= 3cm}]
        \nextgroupplot[
            legend cell align={left},
            legend style={
                    fill opacity=0.8,
                    draw opacity=1,
                    text opacity=1,
                    at={(0.5,0.5)},
                    anchor=center,
                    font=\LARGE,
                    draw=lightgray204
                },
            tick align=outside,
            tick pos=both,
            title={(a) DenseNet-121 and CIFAR-10},
            x grid style={darkgray176},
            xlabel={\# of Faulty Clients},
            xmin=1.75, xmax=7.25,
            xtick style={color=black},
            y grid style={darkgray176},
            label style={font=\huge},
            tick label style={font=\huge},
            title style={font=\huge},
            ylabel={Time (s)},
            ymin=0, ymax=7,
            ytick style={color=black}
        ]
        \addplot [line width=2.4pt, black, mark=*, mark size=3, mark options={solid}, smooth]
        table {%
                2 1.2407248
                3 1.37257292
                5 1.21690741
                7 1.21556752
            };
        \addlegendentry{Input Generation Time}
        \addplot [line width=2.4pt, black, dashed, mark=triangle*, mark size=3, mark options={solid}, smooth]
        table {%
                2 4.943
                3 5.18
                5 5.4
                7 5.802
            };
        \addlegendentry{Fault Localization Time}

        \nextgroupplot[
            legend cell align={left},
            legend style={
                    fill opacity=0.8,
                    draw opacity=1,
                    text opacity=1,
                    at={(0.5,0.8)},
                    anchor=center,
                    font=\LARGE,
                    draw=lightgray204
                },
            tick align=outside,
            tick pos=both,
            label style={font=\huge},
            tick label style={font=\huge},
            title style={font=\huge},
            title={(b) ResNet-50 and CIFAR-10},
            x grid style={darkgray176},
            xlabel={\# of Faulty Clients},
            xmin=1.75, xmax=7.25,
            xtick style={color=black},
            y grid style={darkgray176},
            ylabel={Time (s)},
            ymin=0, ymax=7,
            ytick style={color=black}
        ]
        \addplot [line width=2.4pt, black, mark=*, mark size=3, mark options={solid}, smooth]
        table {%
                2 1.05540812
                3 1.05512111
                5 1.09866641
                7 1.06449928
            };
        \addlegendentry{Input Generation Time}
        \addplot [line width=2.4pt, black, dashed, mark=triangle*, mark size=3, mark options={solid}, smooth]
        table {%
                2 3.567
                3 3.913
                5 4.045
                7 4.224
            };
        \addlegendentry{Fault Localization Time}
    \end{groupplot}

\end{tikzpicture}

%% file: graphs/tikz_scalability_400.tex
\begin{tikzpicture}

    \definecolor{darkgray176}{RGB}{176,176,176}
    \definecolor{lightgray204}{RGB}{204,204,204}

    \begin{groupplot}[group style={group size=3 by 1, horizontal sep = 2.5cm}]
        
        \nextgroupplot[
            tick align=outside,
            tick pos=both,
            label style={font=\huge},
            tick label style={font=\LARGE},
            title style={font=\huge},
            title={(a)},
            x grid style={darkgray176},
            xlabel={Clients},
            xmin=6.25, xmax=418.75,
            xtick style={color=black},
            y grid style={darkgray176},
            ylabel={Localization Accuracy},
            ymin=0, ymax=100,
            ytick style={color=black}
        ]
        \addplot [line width=2.4pt, black, smooth]
        table {%
                25 100
                50 90
                80 90
                200 85
                400 75
            };

                \nextgroupplot[
            legend cell align={left},
            legend style={
                    fill opacity=0.8,
                    draw opacity=1,
                    text opacity=1,
                    font=\LARGE,
                    at={(0.03,0.97)},
                    anchor=north west,
                    draw=lightgray204
                },
            log basis y={10},
            tick align=outside,
            tick pos=both,
            label style={font=\huge},
            tick label style={font=\LARGE},
            title style={font=\huge},
            title={(b)},
            x grid style={darkgray176},
            xlabel={Clients},
            xmin=6.25, xmax=418.75,
            xtick style={color=black},
            y grid style={darkgray176},
            ylabel={Time (s), Log Scale},
            ymin=1.05894479911233, ymax=10000,
            ymode=log,
            ytick style={color=black},
            ytick={0.1,1,10,100,1000,10000},
            yticklabels={
                    \(\displaystyle {10^{-1}}\),
                    \(\displaystyle {10^{0}}\),
                    \(\displaystyle {10^{1}}\),
                    \(\displaystyle {10^{2}}\),
                    \(\displaystyle {10^{3}}\),
                    \(\displaystyle {10^{4}}\)
                }
        ]
        \addplot [line width=2.4pt, black,  smooth]
        table {%
                25 1.53421624
                50 1.33364093
                80 2.08869939
                200 5.92733443
                400 12.6547971
            };
        \addlegendentry{Input Generation Time}
        \addplot [line width=2.4pt, black, dashed,  smooth]
        table {%
                25 1.991
                50 5.094
                80 10.242
                200 40.991
                400 134.385
            };
        \addlegendentry{Fault Localization Time}

        \nextgroupplot[
            tick align=outside,
            tick pos=both,
            label style={font=\huge},
            tick label style={font=\LARGE},
            title style={font=\huge},
            title={(c)},
            x grid style={darkgray176},
            xlabel={Data/Client},
            xtick style={color=black},
            y grid style={darkgray176},
            ylabel={Localization Accuracy},
            ymin=0, ymax=100,
            ytick style={color=black}
        ]
        \addplot [line width=2.4pt, black, smooth]
        table {%
                850 75
                1700 85
                4250 90
                6800 90
                13600 100
            };

    \end{groupplot}

\end{tikzpicture}

%% file: graphs/tikz_nc_thresholds.tex


\begin{tikzpicture}

\definecolor{darkgray176}{RGB}{176,176,176}

\begin{groupplot}[
group style={group size=2 by 2,
vertical sep=3cm,
horizontal sep =3cm,    
}
]

\nextgroupplot[
tick align=outside,
tick pos=both,
title={(a) ResNet-50 and CIFAR-10},
x grid style={darkgray176},
xmin=-0.045, xmax=0.945,
xtick style={color=black},
y grid style={darkgray176},
label style={font=\huge},
tick label style={font=\huge},
title style={font=\huge},
ylabel={Localization Accuracy (\%)},
ymin=53.8, ymax=102.2,
ytick style={color=black}
]
\addplot [line width=2.4pt, black, mark=triangle*, mark size=3, mark options={solid}, smooth]
table {%
0 100
0.001 100
0.003 100
0.01 100
0.05 100
0.1 100
0.3 90
0.5 92
0.7 56
0.9 72
};

\nextgroupplot[
tick align=outside,
tick pos=both,
label style={font=\huge},
tick label style={font=\huge},
title style={font=\huge},
title={(b) ResNet-50 and FEMNIST},
x grid style={darkgray176},
xmin=-0.045, xmax=0.945,
xtick style={color=black},
y grid style={darkgray176},
ymin=5.5, ymax=104.5,
ytick style={color=black}
]
\addplot [line width=2.4pt, black, mark=triangle*, mark size=3, mark options={solid}, smooth]
table {%
0 100
0.001 100
0.003 100
0.01 98
0.05 80
0.1 80
0.3 70
0.5 38
0.7 10
0.9 10
};

\nextgroupplot[
tick align=outside,
tick pos=both,
title={(c) DenseNet-121 and CIFAR-10},
x grid style={darkgray176},
xlabel={Neuron Activation Threshold},
xmin=-0.045, xmax=0.945,
xtick style={color=black},
label style={font=\huge},
tick label style={font=\huge},
title style={font=\huge},
y grid style={darkgray176},
ylabel={Localization Accuracy (\%)},
ymin=3.4, ymax=104.6,
ytick style={color=black}
]
\addplot [line width=2.4pt, black, mark=triangle*, mark size=3, mark options={solid}, smooth]
table {%
0 100
0.001 100
0.003 100
0.01 100
0.05 100
0.1 100
0.3 100
0.5 100
0.7 60
0.9 8
};

\nextgroupplot[
tick align=outside,
tick pos=both,
title={(d) DenseNet-121 and FEMNIST},
x grid style={darkgray176},
xlabel={Neuron Activation Threshold},
xmin=-0.045, xmax=0.945,
xtick style={color=black},
label style={font=\huge},
tick label style={font=\huge},
title style={font=\huge},
y grid style={darkgray176},
ymin=16, ymax=104,
ytick style={color=black}
]
\addplot [line width=2.4pt, black, mark=triangle*, mark size=3, mark options={solid}, smooth]
table {%
0 100
0.001 100
0.003 100
0.01 100
0.05 100
0.1 100
0.3 80
0.5 64
0.7 22
0.9 20
};
\end{groupplot}

\end{tikzpicture}

%% file: paper_sections/section_threat_to_validity.tex
\subsection{Threats to Validity}
\label{section:threats}
To alleviate threats to external validity, we use established state-of-the-art FL experimental models (ResNet-18, ResNet-34, ResNet-50, DenseNet-121, and VGG-16), two standardized datasets from FL benchmarks, two real-world data distributions, and an industrial scale FL framework. Similarly, we remove bias in fault injection using standard noisy labels technique from the ML literature, to make a fault reflective of real-world scenarios. We also experiment with varying noise rates for better evaluations, transparency, and fairness. Another source of external threats to validity is randomness in \tool's input selection method. We minimize such randomness by evaluating each configuration on at least 10 and 100 test inputs and reporting the average results. 





%% file: paper_sections/section_related-work.tex
\section{Related Work}
\label{related-work}

Debugging ML models has been extensively explored in recent works~\cite{pei2017deepxplore, xie2019diffchaser, guo2018dlfuzz, usman2021neurospf, wardat2021deeplocalize, odena2019tensorfuzz, braiek2019deepevolution}. The primary objectives of these approaches are interpretability, generating new test cases by carefully perturbing the real-world training inputs to improve performance and to find bugs and corner cases in the given model. These approaches require access to the training and testing data, and some are limited to testing a single neural network; hence, such approaches cannot be directly imported into FL. Lack of access to client data and resources in FL settings makes testing and debugging FL more challenging. If applied to FL, these testing approaches will find every client's model defective.  Clients' models are architecturally similar but trained on local clients' data, and thus their models are semantically different from each other. Identifying defects in an isolated model is not practical either. Every client's model has weaknesses that will surface on carefully selected test data. \tool overcomes these problems by focusing on the commonality of models instead of differences. 

Most relevant work to {\tool} primarily focuses on finding clients' contributions to a global model without exposing the private data to a central server~\cite{zeng2021comprehensive}. In practice, individual clients report information about training, such as dataset size and performance metrics, to the central aggregator~\cite{kang2019incentive, sarikaya2019motivating, zeng2020fmore, ye2020federated, le2021incentive}. Existing approaches use prior information \eg previous task performance and data quality obtained via third-party services, to evaluate clients' models~\cite{ur2021trustfed}. Other approaches recommend cross-validating clients' models on another client's local dataset~\cite{lyu2020towards}. Another alternate is maintaining a validation dataset at the central server to evaluate clients' models~\cite{lyu2020collaborative, chen2020focus}.  A major limitation of the above FL-related approaches is that the aggregator server depends entirely on the client\textquotesingle s reported information or test data to evaluate clients' models. The aggregator also assumes that all clients are trustworthy about their performance in these approaches, which attracts adversarial clients like the ones in targeted poisoning attacks~\cite{ozdayi2021defending}. 
Cross-validation is also prohibited due to limited computing resources for edge devices such as smart home sensors. \tool overcomes the limitations of debugging faulty clients with interactive and automated approaches that preserve privacy.

%% file: paper_sections/section_conclusion.tex
\section{Conclusion}
\label{conclusion}
Federated learning promotes collaborative model training across millions of clients---the type of learning that was previously impossible due to privacy concerns related to user data. However, FL poses unprecedented challenges in debugging a faulty client responsible for deterring global training. With minimal information about the training process and non-existent debugging techniques, such issues are often left untreated. \tool enables interactive and automated fault localization in FL. It adapts conventional debugging practices in FL with its {\em breakpoint} and {\em fix and replay} feature. It offers a novel differential testing technique to automatically identify the precise faulty clients. We demonstrate that \tool identifies a faulty client with 100\% accuracy within 2.1\% of a round's training time, advocating for \tool's efficacy and efficiency. With \tool, we pave the way for advanced software debugging techniques to be adapted in the emerging area of federated learning and the broader community of machine learning practitioners.

\section*{Acknowledgements}
This work is supported in part by the National Science Foundation under grant number CCF-2106420.
We would like to thank the ICSE reviewers for their constructive feedback. 

\balance